\documentclass[preprint,tightenlines,showpacs,amsmath,amssymb]{revtex4}

\usepackage{epsfig,epsf,graphics,psfrag}
\usepackage{times}
\usepackage{float}
\usepackage{color}
\usepackage{amsfonts,amssymb,stmaryrd,latexsym,amsmath}
\usepackage{multirow}
\usepackage{array}
\usepackage{hhline}
\usepackage{booktabs}
\usepackage{enumerate}
\usepackage{slashed}
\usepackage{subfigure}
\usepackage[colorlinks, citecolor=blue,anchorcolor=red,menucolor=red, linkcolor=red,filecolor=red,runcolor=red,urlcolor=blue,frenchlinks=red]{hyperref}

\begin{document}

\bibliographystyle{unsrt}

\title{Triangle singularity as the origin of $X_0(2900)$ and $X_1(2900)$ observed in $B^+\to D^+ D^- K^+$}

\author{Xiao-Hai~Liu$^{1}$}\email{xiaohai.liu@tju.edu.cn}
\author{Mao-Jun Yan$^{2}$}\email{yanmaojun@buaa.edu.cn}
\author{Hong-Wei Ke$^{1}$}\email{khw020056@tju.edu.cn}
\author{Gang Li$^{3}$}\email{gli@qfnu.edu.cn}
\author{Ju-Jun Xie$^{4,5,6}$}\email{xiejujun@impcas.ac.cn}

\affiliation{
	$^1$Department of Physics, School of Science, Tianjin University, Tianjin 300350, China\\
	$^2$School of Physics,Beihang University, Beijing 100191, China \\
	$^3$College of Physics and Engineering, Qufu Normal University, Qufu 273165, China\\
	$^4$Institute of Modern Physics, Chinese Academy of Sciences, Lanzhou 730000, China\\
	$^5$School of Nuclear Sciences and Technology, University of Chinese Academy of Sciences, Beijing 101408, China\\
    $^6$School of Physics and Microelectronics, Zhengzhou University, Zhengzhou, Henan 450001, China}
\date{\today}

\begin{abstract}

The LHCb collaboration reported the observation of a narrow peak in the $D^- K^+$ invariant mass distributions from the $B^+\to D^+ D^- K^+$ decay. The peak is parameterized in terms of two resonances $X_0(2900)$ and $X_1(2900)$ with the quark contents $\bar{c}\bar{s}ud$, and their spin-parity quantum numbers are $0^+$ and $1^-$, respectively. We investigate the rescattering processes which may contribute to the $B^+\to D^+ D^- K^+$ decays. It is shown that the $D^{*-}K^{*+}$ rescattering via the $\chi_{c1}K^{*+}D^{*-}$ loop or the  $\bar{D}_{1}^{0}K^{0}$ rescattering via the $D_{sJ}^{+}\bar{D}_{1}^{0}K^{0}$ loop simulate the $X_0(2900)$ and $X_1(2900)$ structures. Such phenomena are due to the analytical property of the scattering amplitudes with the triangle singularities located to the vicinity of the physical boundary.



\pacs{~14.40.Rt,~12.39.Mk,~14.40.Nd}

\end{abstract}

\maketitle

\section{Introduction}

The study on exotic hadrons is experiencing a renaissance. Dozens of $XYZ$ particles which do not fit into the conventional quark model predictions are observed since 2003. Most of these $XYZ$ particles are thought to be tetraquark or hadronic molecule states containing a hidden $c\bar{c}$ or $b\bar{b}$. In 2016, the D0 collaboration reported the observation of a narrow state $X(5568)$ in the $B_s^0\pi^\pm$ invariant mass spectrum~\cite{D0:2016mwd}. This $X(5568)$ was then thought to be a fully open-flavor exotic hadron with the quark contents $su\bar{b}\bar{d}$ (or $sd\bar{b}\bar{u}$). However, in a later experimental result reported by the LHCb collaboration \cite{Aaij:2016iev}, the existence of $X(5568)$ was not confirmed based on their $pp$ collision data, and the existence of this state was also severely challenged on theoretical grounds~\cite{Burns:2016gvy,Guo:2016nhb,Kang:2016zmv}. The possible reason of its appearance in the D0 and absence in LHCb and CMS experiments was discussed in Ref.~\cite{Yang:2016sws}. We refer to Refs.~\cite{Brambilla:2019esw,Olsen:2017bmm,Guo:2017jvc,Chen:2016spr,Chen:2016qju} for some recent reviews about the study on $XYZ$ particles.

Very recently, the LHCb collaboration reported the observation of two resonance-like structures in $D^- K^+$ invariant mass distributions in $B^+\to D^+ D^- K^+$ decays. Their masses and widths are
\begin{eqnarray}\label{Xstates}
&&	M_{X_0(2900)}=2.866 \pm 0.007~ \mbox{GeV},\ 	\Gamma_{X_0(2900)}= 57.2\pm 12.9~  \mbox{MeV}, \\
&&	M_{X_1(2900)}=2.904 \pm 0.005~ \mbox{GeV},\ 	\Gamma_{X_1(2900)}=110.3\pm 11.5~ \mbox{MeV}.
\end{eqnarray}
Their spin-parity quantum numbers $J^P$ are $0^+$ and $1^-$, respectively. Since they are observed in the $D^-K^+$ channel, the $X_0(2900)$ and $X_1(2900)$ should be states with four different valence quarks $\bar{c}\bar{s}ud$~\cite{lhcbexptalk}. Before the LHCb observation the tetraquark states with four different flavors have been systematically investigated in Ref.~\cite{Cheng:2020nho} using a color-magnetic interaction model. An excited scalar tetraquark state with the mass 2850 MeV and the quark contents $cs\bar{u}\bar{d}$ is predicted in Ref.~\cite{Cheng:2020nho}, which may account for the $X_0(2900)$. Another state with the mass 2902 MeV and the same quark contents is also predicted, but its spin-parity is $1^+$, which is not consistent with the current experiment. In a very recent paper of Ref.~\cite{Karliner:2020vsi}, the newly observed $X_0(2900)$ is interpreted as a $\bar{c}\bar{s}ud$ isosinglet compact tetarquark state, but the broader $1^-$ peak is not interpreted in the same tetraquark framework.

Concerning the nature of these $XYZ$ states, apart from the genuine resonances interpretations, some non-resonance interpretations were also proposed in literature, such as the cusp effect or the triangle singularity (TS) mechanism. It is shown that sometimes it is not necessary to introduce a genuine resonance to describe a resonance-like peak, because some kinematic singularities of the rescattering amplitudes could behave themselves as bumps in the corresponding invariant mass distributions and simulate genuine resonances, which may bring ambiguities to our understanding about the nature of exotic states.
Before claiming that one resonance-like peak corresponds to one genuine particle, it is also necessary to exclude or confirm these possibilities. We refer to Ref.~\cite{Guo:2019twa} for a recent and detailed review about the threshold cusps and TSs in hadronic reactions.

In this work, we study the $B^+ \to D^+D^-K^+$ reaction by considering some possible triangle rescattering processes, and try to provide a natural explanation for the exotic hadron candidates $X_0(2900)$ and $X_1(2900)$ reported by LHCb.

\section{The Model}

\begin{figure}[htbp]
	\centering
	\includegraphics[width=0.8\hsize]{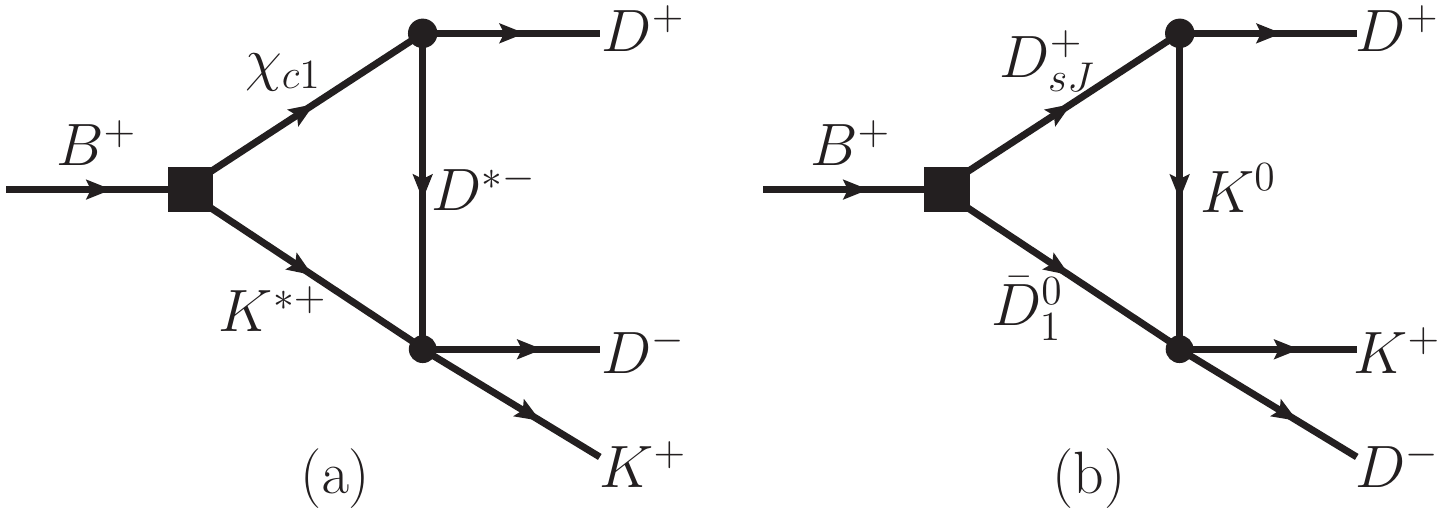}
	\caption{$B^+\to D^+ D^- K^+$ via the triangle rescattering diagrams. Kinematic conventions for the intermediate states are (a) $K^{*+}(q_1)$, $\chi_{c1}(q_2)$, $D^{*-} (q_3)$ and (b) $\bar{D}_1^{0}(q_1)$, $D_{sJ}^+(q_2)$, $K^0 (q_3)$.}\label{Triangle-Diagram}
\end{figure}

The bottom meson decaying into a charmonium and a kaon meson or a charmed-strange meson and an anti-charmed meson are Cabibbo-favored processes. Therefore it is expected that the rescattering processes illustrated in Figs.~\ref{Triangle-Diagram}(a) and (b) may play a role in the decay $B^+\to D^+ D^- K^+$. In Fig.~\ref{Triangle-Diagram}(a), the intermediate state $\chi_{c1}$ indicates any charmonia with the quantum numbers $J^{PC}=1^{++}$. For the states (nearly) above $D\bar{D}^*$ threshold, the $\chi_{c1}$ could be the experimentally observed $\chi_{c1}(3872)$\footnote[1]{The $X(3872)$ is named $\chi_{c1}(3872)$ by the 
Particle Data Group (PDG) according to its quantum numbers~\cite{Tanabashi:2018oca}.}, $\chi_{c1}(4140)$ or $\chi_{c1}(4274)$~\cite{Tanabashi:2018oca}. In Fig.~\ref{Triangle-Diagram}(b), the intermediate state $D_{sJ}^+$ indicates a higher charmed-strange meson that can decay into $D^+ K^0$. The candidates of $D_{sJ}$ could be $D_{s1}^{*}(2700)$ and $D_{s1}^{*}(2860)$, of which the quantum numbers are $J^{P}=1^{-}$. The $\bar{D}_1^0$ in Fig.~\ref{Triangle-Diagram}(b) represents the narrow anti-charmed meson $\bar{D}_1(2420)^0$ with $J^P=1^+$. 
The thresholds of $D^{*-}K^{*+}$ and $\bar{D}_{1}^{0}K^{0}$ are 2902 and 2918 MeV, respectively. These two thresholds are very close to $M_{X_0(2900)}$ and $M_{X_1(2900)}$. It is therefore natural to expect that the $D^{*-}K^{*+}$ and/or $\bar{D}_{1}^{0}K^{0}$ rescattering and the resulting threshold cusps may account for the LHCb observations.

Another intriguing character of the rescattering processes in Fig.~\ref{Triangle-Diagram} is that the $\chi_{c1}K^{*+}$ and $D_{sJ}^{+}\bar{D}_{1}^0$ thresholds are close to the mass of $B^+$ meson, therefore the TSs of rescattering amplitudes in the complex energy plane could be close to the physical boundary, and the TS may enhance the two-body threshold cusp effect or itself may generate a peak in $D^-K^+$ spectrum. Numerically, if we ignore the widths of intermediate states, when $4311\leq m_{\chi_{c1}} \leq 4388\ \mbox{MeV}$, the TS of the $\chi_{c1}K^{*+}D^{*-}$ loop lies on the physical boundary. The observed $\chi_{c1}(4274)$ is very close to this region. For the $D_{sJ}^{+}\bar{D}_{1}^{0}K^{0}$ loop, when $2539\leq m_{D_{sJ}} \leq 2858\ \mbox{MeV}$, the TS lies on the physical boundary~\cite{Guo:2019twa}. There are several charmed-strange mesons whose masses fall in this kinematic range, such as $D_{s1}^{*}(2700)$ and $D_{s1}^{*}(2860)$.

Since $M_{B^+}$ is close to the $\chi_{c1}K^{*+}$ and $D_{sJ}^{+}\bar{D}_{1}^{0}$ thresholds, the $S$-wave decays are expected to be dominated. The general $S$-wave decay amplitudes can be written as
\begin{eqnarray}\label{Bplusdecays}
\mathcal{A}(B^+\to \chi_{c1}K^{*+}) &=& g_{W}^a \epsilon^*_{\chi_{c1}}\cdot \epsilon^*_{{K}^{*+}}, \\ 
\mathcal{A}(B^+\to D_{sJ}^{+}\bar{D}_{1}^{0}) &=& g_{W}^b \epsilon^*_{D_{sJ}^{+}}\cdot \epsilon^*_{\bar{D}_{1}^{0}},
\end{eqnarray}
where $g_{W}^a$ and $g_{W}^b$ represent the weak couplings, and we take the quantum numbers of $D_{sJ}^{+}$ to be $J^P=1^-$.

For the processes $\chi_{c1}\to D^+ D^{*-}$ and $D_{sJ}^{+}\to D^+ K^0$, the amplitudes read
\begin{eqnarray}\label{vertex2}
\mathcal{A}(\chi_{c1}\to D^+ D^{*-}) &=& g_{\chi_{c1} D \bar{D}^{*}} \epsilon_{\chi_{c1}}\cdot \epsilon^*_{{D}^{*-}}, \\ 
\mathcal{A}(D_{sJ}^{+}\to D^+ K^0) &=& g_{D_{sJ}DK} (p_{D^+}-p_{K^0})\cdot \epsilon_{D_{sJ}^{+}}.
\end{eqnarray}

The quantum numbers of  $D^- K^+$ system in relative $S$-, $P$-, and $D$-wave are $J^P=0^+$, $1^-$ and $2^+$, respectively. For the rescattering processes in Figs.~\ref{Triangle-Diagram}(a) and (b), we are interested in the near-threshold $S$-wave $D^{*-} K^{*+}$ (or $\bar{D}_{1}^{0} K^{0}$) scattering into $D^- K^+$. By taking into account requirements of the parity (strong interaction vertices) and angular momentum conservation, the quantum numbers of $D^- K^+$ system are $0^+$ and $1^-$ for Figs.~\ref{Triangle-Diagram}(a) and (b), respectively. The scattering amplitudes for $D^{*-} K^{*+} \to D^- K^+$ and $\bar{D}_{1}^{0} K^{0} \to D^- K^+$ can be written as
\begin{eqnarray}\label{vertex3}
\mathcal{A}(D^{*-} K^{*+} \to D^- K^+) &=& C_{a} \epsilon_{D^{*-}}\cdot \epsilon_{K^{*+}}, \\ 
\mathcal{A}(\bar{D}_{1}^{0} K^{0} \to D^- K^+) &=& C_{b} (p_{D^-}-p_{K^+})\cdot \epsilon_{\bar{D}_{1}^{0}},
\end{eqnarray}
where $C_a$ and $C_b$ are constants.

The decay amplitude of $B^+\to D^+ D^- K^+$ via the $\chi_{c1}K^{*+}D^{*-}$ loop in Fig.~\ref{Triangle-Diagram} (a) is given by
\begin{eqnarray}\label{loop-amplitude-a}
&&\mathcal{A}_{B^+\to D^+ D^- K^+}^{[ \chi_{c1}K^{*+}D^{*-} ]} = -{i} \int \frac{d^4q_1}{(2\pi)^4} \frac{\mathcal{A}(B^+\to \chi_{c1}K^{*+})  }{ (q_1^2-m_{{K}^*}^2 +i m_{{K}^*}\Gamma_{{K}^*})  }  \nonumber \\
&&\times \frac{ \mathcal{A}(\chi_{c1}\to D^+ D^{*-}) \mathcal{A}(D^{*-} K^{*+} \to D^- K^+) }{ (q_2^2-m_{\chi_{c1}}^2 +i m_{\chi_{c1}}\Gamma_{\chi_{c1}}) (q_3^2-m_{D^{*}}^2) } ,
\end{eqnarray}
where the sum over polarizations of intermediate state is implicit. The amplitude of Fig.~\ref{Triangle-Diagram}(b) is similar to that of Fig.~\ref{Triangle-Diagram}(a) and reads
\begin{eqnarray}\label{loop-amplitude-b}
&&\mathcal{A}_{B^+\to D^+ D^- K^+}^{[ D_{sJ}^{+}\bar{D}_{1}^{0}K^{0} ]} = -{i} \int \frac{d^4q_1}{(2\pi)^4} \frac{\mathcal{A}(B^+\to D_{sJ}^{+}\bar{D}_{1}^{0})  }{ (q_1^2-m_{\bar{D}_{1}}^2 +i m_{\bar{D}_{1}}\Gamma_{\bar{D}_{1}})  }  \nonumber \\
&&\times \frac{ \mathcal{A}(D_{sJ}^{+}\to D^+ K^0) \mathcal{A}(\bar{D}_{1}^{0} K^{0} \to D^- K^+) }{ (q_2^2-m_{D_{sJ}}^2 +i m_{D_{sJ}}\Gamma_{D_{sJ}}) (q_3^2-m_{K}^2) }.
\end{eqnarray}
As long as the TS kinematical conditions are satisfied, it implies that one of the intermediate state (here $\chi_{c1}$ for Fig.~\ref{Triangle-Diagram}(a) and $D_{sJ}^+$ for Fig.~\ref{Triangle-Diagram}(b) respectively) must be unstable. It is therefore necessary to take into account the width effect. We use the Breit-Wigner (BW) type propagators to account for the width effects of intermediate states, or equivalently replace the real
mass $m$ by the complex mass
$m-i\Gamma/2$~\cite{Aitchison:1964zz,Guo:2019twa}. The complex mass in the propagator can remove the TS from physical boundary and makes the physical scattering amplitude finite. Besides, the width effects of ${K}^{*+}$ and $\bar{D}_1^0$ are also taken into account in Eqs.~(\ref{loop-amplitude-a}) and (\ref{loop-amplitude-b}) by employing the BW propagators.

\section{Numerical Results}

\begin{figure}[htbp]
	\centering
	\includegraphics[scale=0.55]{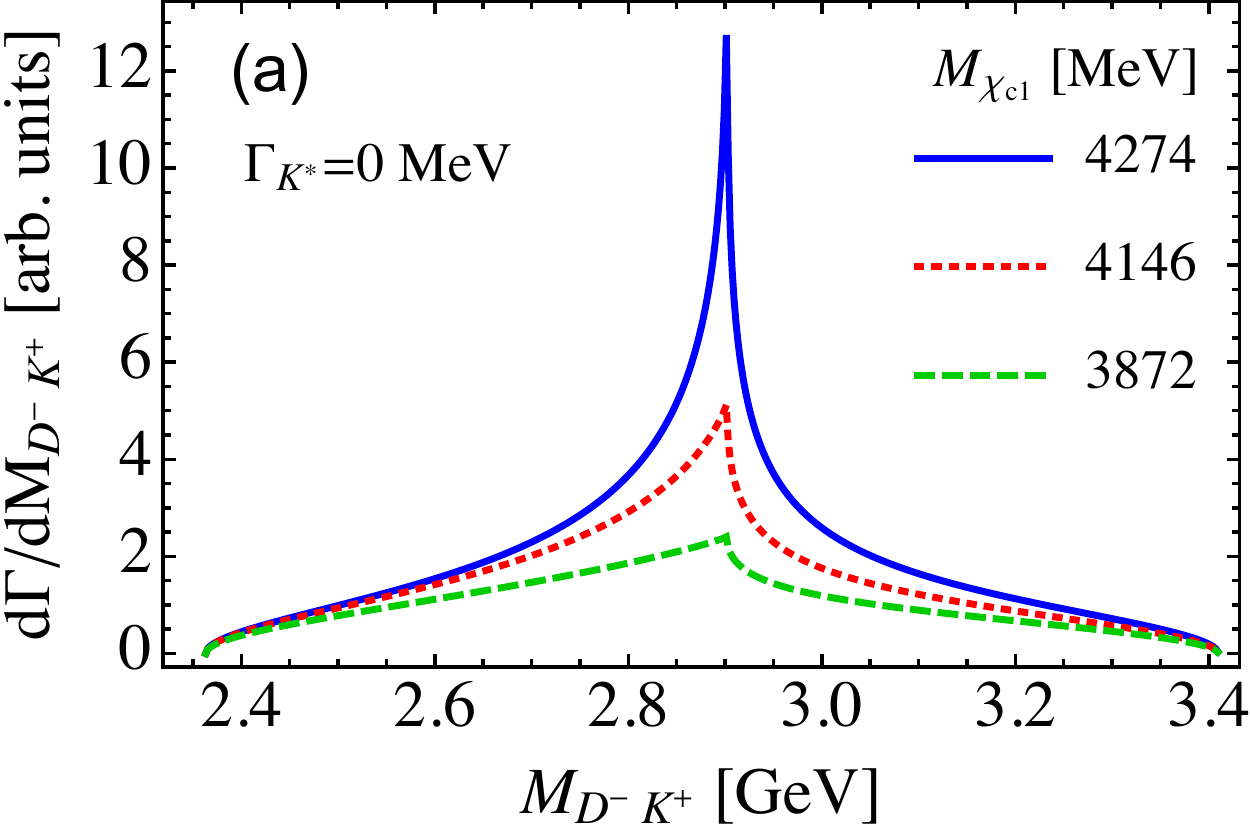} \hspace{0.3cm}
	\includegraphics[scale=0.55]{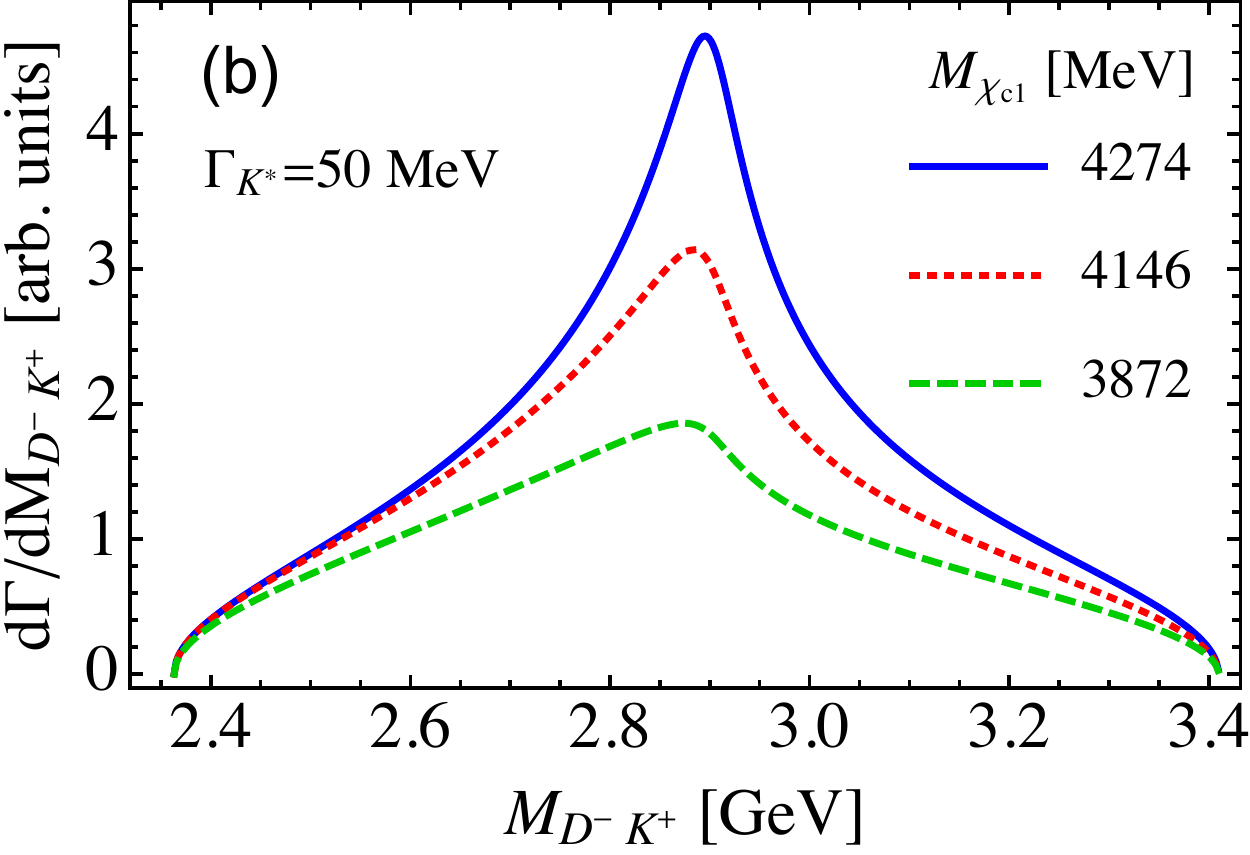}
	\caption{Invariant mass distribution of $D^- K^+$ via the rescattering processes in Fig.~\ref{Triangle-Diagram}(a). The mass and width of intermediate state $\chi_{c1}$ are taken to be those of $\chi_{c1}(4274)$ (solid line), $\chi_{c1}(4140)$ (dotted line), and $\chi_{c1}(3872)$ (dashed line) given by PDG~\cite{Tanabashi:2018oca}, separately. The width of intermediate state $K^{*+}$ is set to be (a) 0 MeV and (b) 50 MeV, respectively. }\label{lineshape-Kstarloop}
\end{figure}

\begin{figure}[htbp]
	\centering
	\includegraphics[scale=0.55]{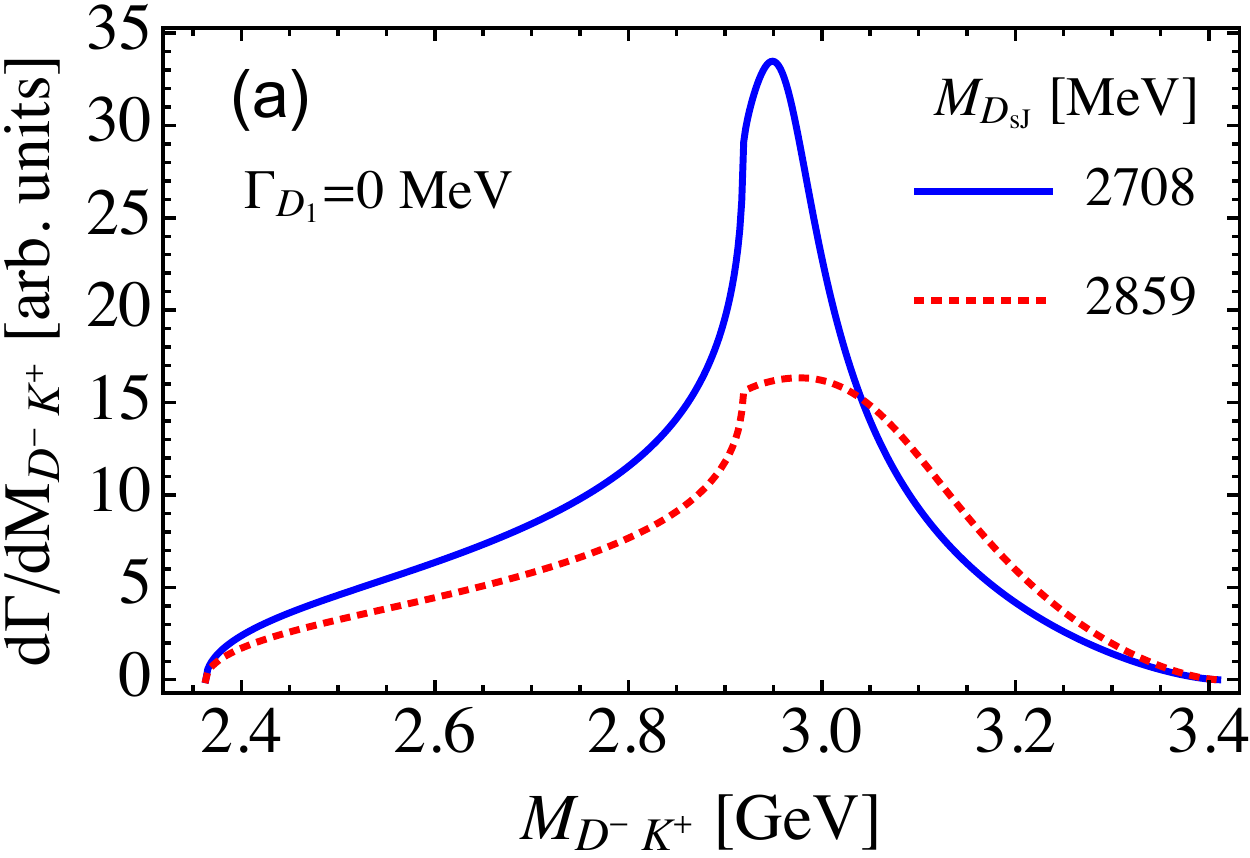} \hspace{0.3cm}
	\includegraphics[scale=0.55]{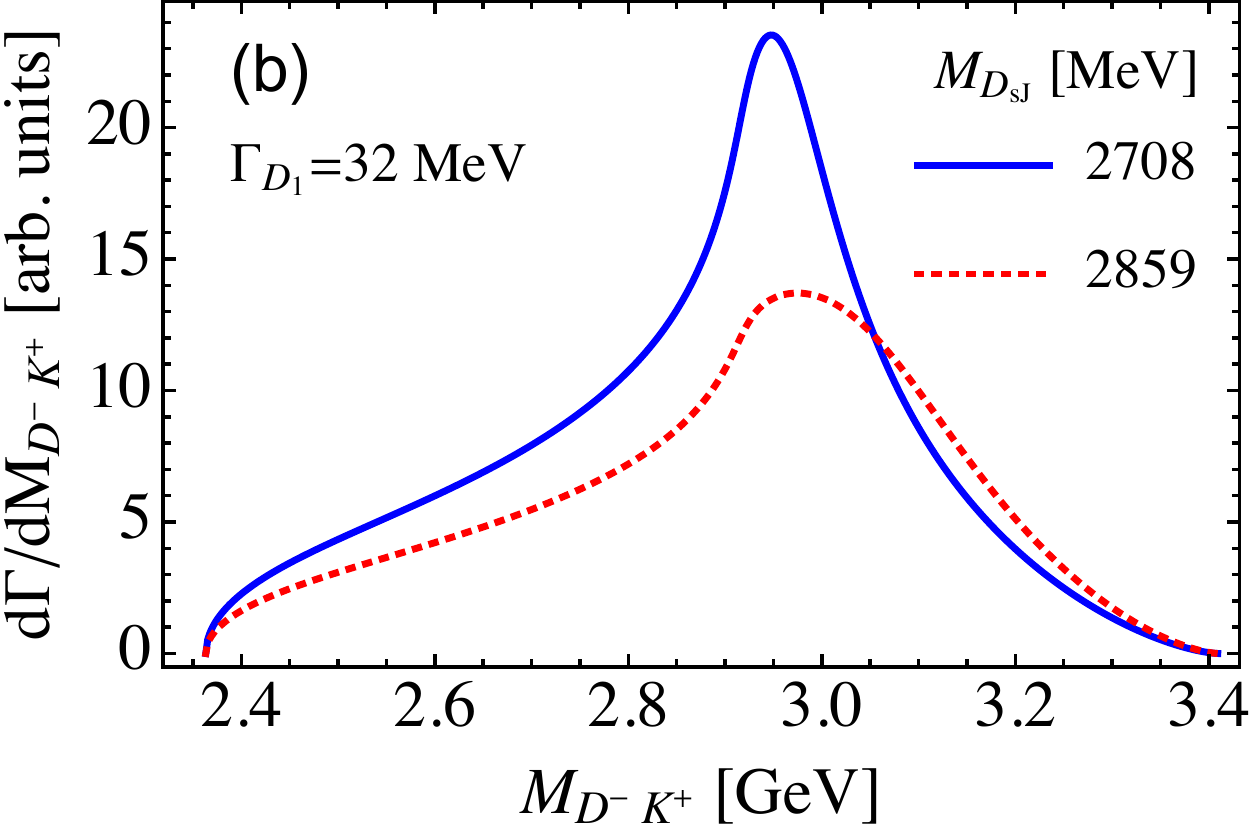}
	\caption{Invariant mass distribution of $D^- K^+$ via the rescattering processes in Fig.~\ref{Triangle-Diagram}(b). The mass and width of intermediate state $D_{sJ}$ are taken to be those of $D_{s1}^{*}(2700)$ (solid line) and $D_{s1}^{*}(2860)$ (dotted line) given by PDG~\cite{Tanabashi:2018oca}, separately. The width of intermediate state $\bar{D}_1(2420)^0$ is set to be (a) 0 MeV and (b) 32 MeV, respectively. }\label{lineshape-D1loop}
\end{figure}

The theoretical results of the $D^- K^+$ invariant mass distributions via the rescattering processes of Figs.~\ref{Triangle-Diagram}(a) and (b) are displayed in Fig.~\ref{lineshape-Kstarloop} and Fig.~\ref{lineshape-D1loop}, respectively. If we ignore the $K^*$ width, it can be seen in Fig.~\ref{lineshape-Kstarloop}(a) there is a sharp peak when the mass of $\chi_{c1}$ is taken to be $m_{\chi_{c1}(4274)}$. This is because $m_{\chi_{c1}(4274)}$ is in the vicinity of the kinematic region where the TS can be present on the physical boundary. When $m_{\chi_{c1}}$ becomes smaller, the TS goes further away from the physical boundary, then its influence becomes insignificant and the relevant peak turns to be broader. If we take into account the $K^*$ width effect, all of the three curves are smoothed to some extent, as illustrated in Fig.~\ref{lineshape-Kstarloop}(b). But the solid line is still comparable with the experimental distribution curve. 

The $\bar{D}_1(2420)$ is relatively narrower compared with $K^*$, therefore the influence of $\bar{D}_1$ width on the line-shape is not quite large, as can be seen from Figs.~\ref{lineshape-D1loop}(a) and (b). When the mass of $D_{sJ}$ is taken to be $m_{D_{s1}^*(2700)}$, a narrow peak around $\bar{D}_1 K$ threshold is obtained, as illustrated with the solid curves in Fig.~\ref{lineshape-D1loop}. We input the center value of $m_{D_{s1}^*(2860)}$ given by PDG, i.e. 2859 MeV, in calculating the dotted curves. Although this mass is just slightly above the kinematic region where the TS can be present on the physical boundary, the mass of $D_{s1}^*(2860)$ is as large as 159 MeV, and this broad width leads to a broad bump in $D^-K^+$ spectrum. Note that there are still some undetermined couplings of the reactions under discussion, we focus on the line-shape of invariant mass distribution curves in this work, which are independent on the values of these coupling constants.

From the invariant mass distribution curve line-shape point of view, it can be seen the rescattering effects can simulate the resonance-like structures $X_0(2900)$ and $X_1(2900)$. Furthermore, from the discussion in Section II, we know the spin-parity quantum numbers of the two resonance-like structures induced by the $D^{*-} K^{*+}$ and $\bar{D}_{1}^{0} K^{0}$ rescatterings are $0^+$ and $1^-$, respectively.
This is also consistent with the current experimental results.

\section{Summary}

In summary, we investigate the $B^+ \to D^+ D^- K^+$ decay via the $\chi_{c1}K^{*+}D^{*-}$ and the $D_{sJ}^{+}\bar{D}_{1}^{0}K^{0}$ rescattering diagrams. Two resonance-like peaks around the $D^{*-}K^{*+}$ and $\bar{D}_{1}^{0}K^{0}$ thresholds are obtained in the $D^-K^+$ invariant mass spectrum. Without introducing genuine exotic states, the two rescattering peaks may simulate the $X_0(2900)$ and $X_1(2900)$ states reported by the LHCb collaboration with consistent spin-parity quantum numbers. 
Such a special phenomenon is due to the
analytical property of the decaying amplitudes with the TS located to the vicinity of the physical boundary. This will enhance the
two-body $D^{*-}K^{*+} \to D^- K^+$ and $\bar{D}^0_1K^0 \to D^-K^+$ rescatterings and make the peak visible in the $D^-K^+$ invariant mass distributions. 

\begin{acknowledgments}
Helpful discussions with Yan-Rui Liu are gratefully acknowledged. This work is supported, in part, by the National Natural Science Foundation of China (NSFC) under Grant Nos.~11975165, 11735003, 11961141012, 11675091, 11835015, and 11375128.
It is also partly supported by the Youth Innovation Promotion Association CAS (No. 2016367).

\end{acknowledgments}

\end{document}